# An Enhanced Buffer Management Scheme for Multimedia Traffic in HSDPA

Suleiman Y. Yerima, Khalid Al-Begain
*Integrated Communications Research Centre*
*Faculty of Advanced Technology, University of Glamorgan*
*Pontypridd (Cardiff) CF37 1DL, Wales, UK*
*E-mail: {syerima,kbegain}@glam.ac.uk*

## Abstract

*High Speed Downlink Packet Access (HSDPA) enables higher data rate packet switch services over UMTS, creating opportunity to provide a variety of broadband multimedia applications. In HSDPA, buffering of downlink data in the base station (Node B) is required to fulfill the Medium Access Control (MAC) Packet Scheduling functionality. Additionally, flow control over the Iub interface that connects the Node B to the Radio Network Controller (RNC) is standardized in the 3GPP HSDPA technical specifications. Thus, buffer queue management schemes can be conveniently applied to improve downlink traffic QoS performance in HSDPA. Our previous work focused on a novel priority queuing scheme shown to be effective for QoS management of multimedia traffic with concurrent diverse flows towards a HSDPA end user. This paper presents an enhanced scheme that incorporates Iub flow control. A dynamic HSDPA simulator is developed to study the extended scheme, demonstrating the performance improvement achievable.*

## 1. Introduction

Third generation (3G) Universal Mobile Telecommunication System (UMTS) was introduced to support higher data rate applications unavailable in previous cellular generations. HSDPA is a technology specified by the 3G Partnership Project (3GPP) to enhance UMTS Radio Access Networks (UTRAN) capacity to support broadband services like multimedia conferencing, VoIP, or high-speed internet access. The ability to support high data rates will enable application developers to create content rich 'multimedia' applications, typically consisting of a number of classes of media or data- with different Quality of Service (QoS) requirements- being concurrently downloaded to a single user [1]. HSDPA significantly reduces downlink transmission latency, enabling peak data rates of up to 14.4 Mbps in addition to a three-fold capacity increase in UMTS networks [2], [3]. A shared downlink channel is utilized, which adapts transmission capacity to changing radio propagation conditions (fast link adaptation). Fast link adaptation employs adaptive modulation and coding (AMC) whereby different modulation and coding schemes are selected for transmission of traffic to the User Equipments (UE) within a serving HSDPA cell. AMC scheme selection is based on the experienced radio channel quality of the UE. Other features of HSDPA include HARQ for error control, and channel-dependent Fast Scheduling.

Figure 1 shows the entities in a HSDPA Radio Access Network. The base station or Node B is responsible for scheduling packets to the UEs within a cell, unlike in basic UMTS where it is handled by the Radio Network Controller (RNC). The inclusion of packet scheduling in the Node B presents opportunity to implement buffer queue management in the Node B to improve QoS guarantees for streams of traffic comprising diverse flows or 'multimedia' traffic.

Buffer management for QoS control in HSDPA Node Bs have been studied in our previous work [4]. In particular, a combined Time-Space Priority (TSP) buffer management strategy for 'multimedia' traffic QoS control over HSDPA downlink in Node B buffers was investigated. This paper explores the extension of the TSP scheme to incorporate a possible threshold-based flow control mechanism applied between the RNC and Node B i.e. over the **Iub** interface that connects the two entities.

The organization of the rest of the paper is as follows. Section 2 reviews related work, while the Node B buffer management schemes are presented in section 3. The HSDPA simulation model for performance study is discussed in section 4, followed by results and analyses in section 5. Finally, the paper ends with some concluding remarks and highlights of further investigation to undertake.

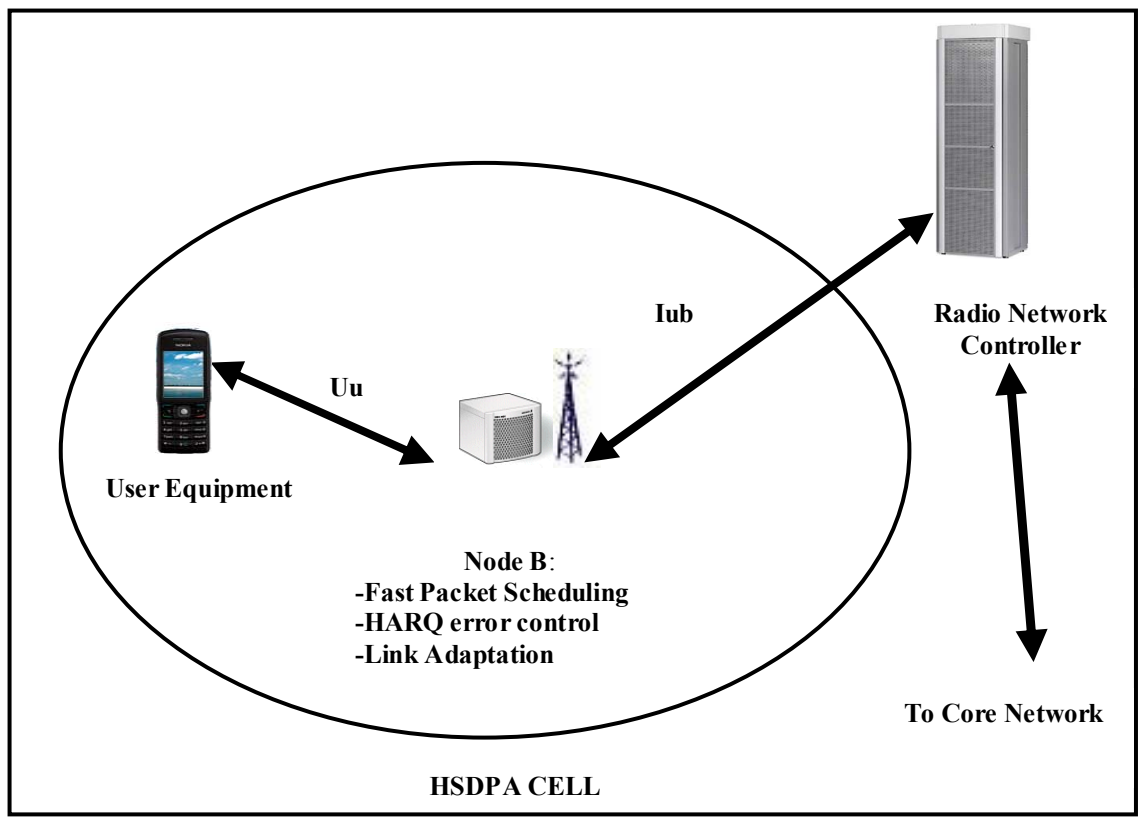

**Figure 1. HSDPA Radio Access Network with additional Node B enhancing functions**

# 2. Related work

Threshold-based queue management techniques have been extensively studied in the literature. In [5] an analytical queuing model for a discrete-time finite queue incorporating two thresholds used to control the sending rate of packets from the traffic source is derived. The system is proposed as a model for congestion control but considers only one traffic class. In [6], an adaptive buffer management scheme for buffers aggregating traffic from various sources with prioritization based on a set of adaptive thresholds is presented. The analysis proves the scheme's capability to provide QoS to higher priority traffic classes, but no flow or rate control is applied.

Active Queue Management (AQM) is a threshold-based queue management technique which is used to control the number of packets in a queue. It operates by dropping packets when necessary, to manage the length of a queue. Random Early Drop (RED) [7] is a well-known AQM algorithm recommended by Internet Engineering Task force (IETF) for Internet routers as a congestion control mechanism and to replace the traditional tail drop queuing management. A novel analytical model for a finite queuing system with AQM under two heterogeneous classes of traffic is reported in [8]. The model employs two thresholds which control the dropping rate of queued packets the queue to ease congestion. Similar to the work in [8], the study in [9] describes a simulation study of Extended RED mechanism for a finite queue of two classes of traffic, but with two sets of thresholds considered for each class of traffic. In [10], a threshold-based queue management is applied for QoS-aware rate control and uplink bandwidth allocation for polling services in IEEE 802.16 wireless networks.

The work in [11] investigates the impact of Iub flow control on HSDPA performance, while an optimized Iub flow control strategy is proposed in [12]. The findings in these works and other similar works suggest the importance of Iub flow control strategies as a key element in HSDPA performance. The approach in this paper is to jointly consider Iub flow control and intra-user priority queuing in a buffer management scheme, that is designed for QoS control of concurrent diverse flows downloaded during a HSDPA user's multimedia session. The scheme is applied in the Node B where MAC-hs data buffers queue packet data units (PDUs) arriving from the RNC over the Iub interface.

# 3. HSDPA buffer management

## 3.1. Time-Space Priority queuing

The Time-Space Priority (TSP) queue management scheme [13], combines time priority and space priority schemes with fixed or variable thresholds to control the QoS parameters (loss, delay, and jitter) of diverse flows comprising a multimedia stream in the same session. Thus, real time (RT) flows, e.g. generated from Video or Voice packets, are given service priority because of their stringent delay requirements; while non real time (NRT) flows, such as email or File downloads, have buffer space priority to minimize loss. This concept is illustrated in Fig. 2. When applied to a user data buffer (consisting of a single queue) in the Node B, arriving PDUs generated from RT packets will be queued in front of the NRT PDUs to receive non-pre-emptive priority scheduling for transmission on the shared channel. NRT PDUs will only be transmitted when no RT PDUs are present in the buffer. This way, the RT QoS delay requirements would not be compromised. In order to fulfil the QoS of the loss sensitive NRT flow, the number of admitted RT packets is restricted to devote more space to the NRT flow. The TSP threshold limits the number of admitted packets to a maximum ***R***. Setting this threshold also improves the jitter performance of the RT flow.

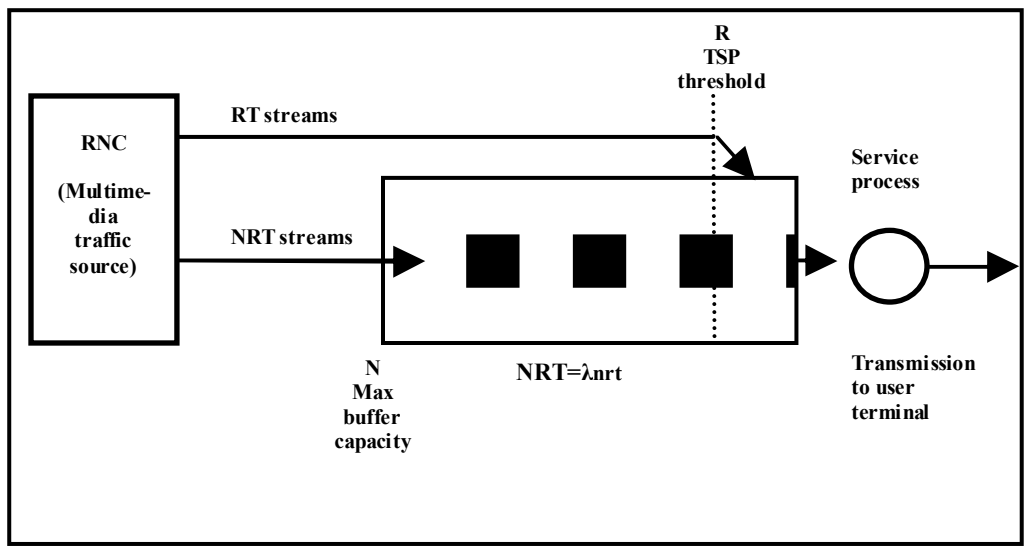

**Figure 2. Original TSP scheme in Node B for a single end user multimedia session without RNC-Node B (Iub) flow control**

### 3.2. HSDPA Iub flow control

In the current HSDPA architecture the MAC layer is split between the RNC and the Node B, and is denoted **MAC-d** and **MAC-hs** in the two entities respectively. On arrival of packets to the RNC, the RLC generates (typically 320 bit) RLC MAC-d service data units (SDU) by segmentation of the packets. From the SDUs MAC-d Protocol data units (PDU) are generated by adding 16 bit header fields resulting in 336 bit PDUs. The MAC-d (Protocol Data Units) PDUs are aggregated and sent over the Iub interface in HS-DSCH Data Frames. A flow control protocol manages the transmission of PDUs over the Iub, which are then buffered in the Node B ready for transmission to the end user in the HSDPA cell. The Node B controls the flow control process by allocating resource grants or credits to the RNC in a signalling message. This signalling message determines the number of MAC-d PDUs that the RNC is allowed to send as well as the time interval over which the PDUs can be sent. See [14] for a detailed explanation of the Iub process.

The flow control signalling protocol is flexible. The RNC can request capacity to send pending PDUs in its buffer. The Node B may respond by granting resource credits to the RNC. Alternatively, credit allocation by the Node B may be unsolicited. The unsolicited mode is particularly useful during periods when excess buffer capacity is available in Node B while the number of pending PDUs in the RNC is high. By using the unsolicited credit allocation mode, buffer underflow, which may lead to underutilization of air interface resources, is prevented.

### 3.3. Enhanced TSP scheme with flow control

In the initially proposed TSP illustrated in figure 2, a single threshold sets the upper bound for maximum allowable RT PDUs in the Node B buffer queue allocated to a UE receiving multimedia traffic over HSDSCH. Drop-tail queue management technique is used to drop arriving NRT PDUs when the buffer queue is full. A drop from tail strategy is also used in a *push out* manner to drop NRT PDUs and accommodate arriving RT PDUs that encounter a full buffer while the RT upper bound is not yet reached. In the enhanced TSP, additional thresholds are included to yield a queue management scheme that controls the number of NRT PDUs requested by the Node B in the signalling message. A possible scheme is shown in figure 3.

The initially allocated PDU arrival rate is given by $\lambda_{nrt}$. When the average number of NRT PDUs queued in the UE's buffer exceeds the lower threshold ***L***, the NRT PDU arrival rate is reduced to $\lambda_q$ , given by:

$$\lambda_q = C \,.\, \lambda_{nrt} \qquad \text{where } 0 < C < 1 \qquad (1)$$

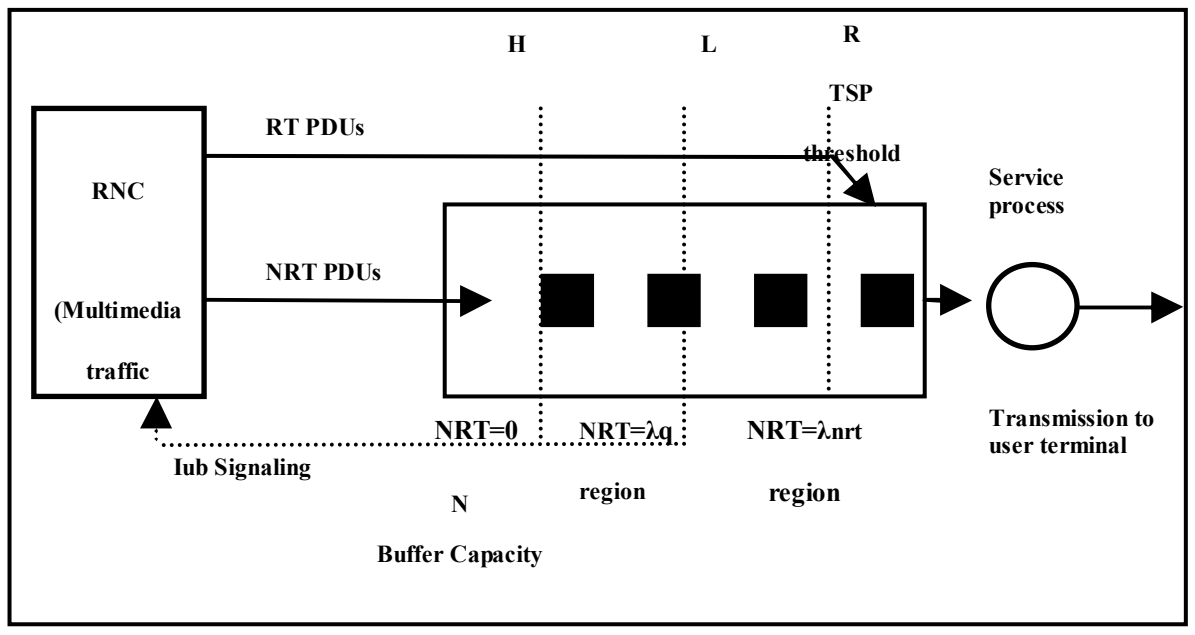

**Figure 3. Enhanced TSP scheme with NRT PDU arrival rate controlled by two thresholds**

PDU arrival rate is reduced by decreasing the maximum credit grants (in PDUs) allocated to the RNC in the next signalling message sent by the Node B. when the average number of queued PDUs exceeds the upper threshold ***H***, the Node B decreases the credit grants to zero to stop the RNC from sending additional PDUs to the UE's buffer in the Node B. Threshold ***H***, is set to less than the maximum buffer capacity ***N***, to allow for spare capacity to absorb instantaneous burst of PDUs that may arrive even after Node B has issued a signalling message to stop further PDU arrivals. This could happen because of the Iub latency (10 to 30ms in practice [11]) which results in a time lag between the instance the signalling message is issued and when it actually arrives at the RNC to update the grant allocation. This spare capacity also eliminates the *push out* dropping policy implemented in the original *drop-tail* TSP scheme which drops NRT PDUs in favour of arriving RT PDUs when the buffer is full while the TSP threshold (***R***) that limits queued RT PDUs is not yet exceeded.

Furthermore, because Iub latency could be much higher than the scheduling interval (if say transmission is scheduled at TTI of 2ms to a UE), queue conditions may change considerably between the instance RNC sends the PDUs and when they finally reach the Node B. Hence, instantaneous queue length is not used as a measure of buffer occupancy. Instead, the average queue length, $Aveq_t$ , is calculated and compared to the thresholds to determine amount of resource grants to issue per HS-DSCH frame. $Ave_q$ is calculated using an exponentially weighted moving average of the queue size which is updated every TTI. $Aveq_t$ is given by:

$$Aveq_t = w_q \,.\, Q_{TTI} + (1 - w_q) \,.\, Aveq_{t-1} \qquad (2)$$

Where $Q_{TTI}$ is the instantaneous queue size at the end of every scheduling transmission time interval (TTI), which is 2ms for HSDPA. Note that $Q_{TTI}$ includes both queued RT and NRT PDUs. Hence, maximum resource grant, G, per HS-DSCH frame is calculated from:

$$G = (\lambda / \text{PDU size}) \cdot TTI_{RLC} \quad (3)$$

Where $TTI_{RLC}$ is the HS-DSCH frame length or frame transmission interval and $\lambda$ is the instantaneous PDU arrival rate to the Node B buffer. $\lambda$ can assume the values $\lambda_{nrt}$ , $\lambda_q$ , and zero depending on the value of $Aveq_t$.

The arrivals of RT PDUs are not rate-controlled by Node B resource credit grants in the manner applicable to the NRT PDUs. Instead, RT PDUs are transferred to the Node B immediately they become available in the RNC MAC-d buffers. The TSP threshold ***R***, like in the original scheme, still sets the upper limit for RT PDUs which ultimately controls RT delay and jitter performance. Even though the RNC is the slave of the Iub flow control while the Node B is the master, the immediate transfer of RT PDUs may be effected using pre-emptive credit allocation [12]. Alternatively the Node B may be configured to grant a fixed maximum credit allocation for RT PDUs equal to the upper bound specified by the TSP threshold, ***R***.

## 4. HSDPA simulation Model

A system level HSDPA simulation model was developed in C++ to evaluate performance of the enhanced TSP scheme. Since the main aim is to explore the viability of the scheme rather than derive exact dimensioning for practical deployment, some simplifications are introduced with only relevant aspects simulated rather than an entire HSDPA UMTS network. The modeled aspects are as follows.

For the traffic model we assumed a multimedia session of concurrent real-time VoIP flows and non-real-time file transfer using FTP. The VoIP traffic is modeled as similar to the model used in [15] with the assumption of ON/OFF periods of 50% probability. The duration of ON/OFF periods is negative exponentially distributed with mean duration of 3 seconds. The VoIP packet length is 304 bits or 38 bytes including RTP/UDP/IP and RLC header overheads. We assumed the ETSI WWW model, with one packet call, for the FTP flow within the concurrent multimedia traffic. Thus FTP average packet size is assumed to be 480 bytes with geometrically distributed inter-arrival times, corresponding to the download bit rate allocated during the multimedia session.

The core network (CN) was not explicitly simulated but assumed to contribute a fixed delay, to the arriving packets without any loss of packets. For the Radio Access Network, a single HSDPA cell served by a Node B under the control of an RNC with the concurrent VoIP and FTP downlink traffic towards an end user (EU) was modelled. Radio propagation was modeled using ITU R pedestrian test path loss model and lognormal shadowing with std. deviation $\sigma = 8$ dB and $\rho = 0.5$. Total Node B transmission power is limited to 15 W, while the allocated HS-PDSCH and CPICH were 7W and 2W respectively, leaving 6 W reserved for other common and dedicated channels.

The instantaneous throughput towards the UE is determined by the transport format (TF) selected, corresponding to a combination of transport block size (TBS), modulation scheme and channel coding rate. This functionality is known as Adaptive modulation and coding (AMC). We assumed that six AMC schemes are available and selected based on the reported channel quality (the instantaneous Signal-to-Interference Ratio, SINR) of the UE under consideration. Due to channel quality transmission latency packets may be received in error as it is possible for the UE SINR at that instance to be different from the last known SINR in the Node B used for AMC selection. Thus HARQ re-transmission is modeled with soft combining of all received packets, the effective SINR being $\boldsymbol{N. SNIR_{init}}$ where N is the number of transmissions and $\boldsymbol{SINR_{init}}$ is the SINR of the first transmission.

Furthermore, the Node B grant allocation interval is chosen to be equal to the sum of the Iub signalling latency, PDU transfer latency and HS-DSCH frame length. MAC-d PDU size length after RLC packet segmentation is taken as 336 bits for both VoIP and FTP flows in the multimedia traffic. Table 1 summarizes the main simulation parameters assumed in the study.

**Table 1. HSDPA simulation parameters**

| Traffic model | |
|---|---|
| VoIP | Packet length=304bits, constant bit rate=15.2 kbps |
| FTP | ETSI WWW model: average packet size= 480 bytes, Geometric inter-arrival times |
| Radio parameters | |
| HS-PDSCH TTI | 2ms |
| Path loss Model | 148 + 40 log (R) dB |
| Transmit powers | Total Node B power=15W, HSDPSCH power=7W, CPICH power=2W |
| Shadow fading | Log-normal: $\sigma = 8$ dB and $\rho = 0.5$ |
| Mobility model | Velocity 3Km/h |
| AMC schemes | QPSK ¼, QPSK ½, QPSK ¾, 16QAM ¼, 16 QAM ½ with multicode transmission |
| CQI latency | 3 HS-PDSCH TTIs (6ms) |
| Cell radius | 1000m |
| RLC parameters | |
| PDU size | 336 bits |
| Iub latency | 20ms |
| HS-DSCH frame | 10ms |

# 5. Numerical results and discussion

In this section we present results of experiments undertaken to compare the performance of the original TSP scheme with the extended one. The HSDPA simulation model described in section 4 is used for the experiments. The performance metrics considered are PDU loss probabilities for both VoIP and FTP flows, mean PDU queuing delay for VoIP and throughput for FTP flows in the multimedia session. The UE receiving the multimedia traffic is assumed to be moving away linearly from the base station (Node B) starting from a position 600m from the cell centre. It was assumed that a maximum of 2 out of the possible 15 HSDPCH channelization codes were available for scheduling the UE's transmission. Performance measures are taken at FTP data rates of $\lambda_{nrt} =$ 64, 128, 256, 512 and 1024 kbps respectively. The buffer management parameters assumed are: ***R***=840 bytes, ***L***=5040 bytes, ***H***= 10,080 bytes, ***N*** =12,600 bytes. Flow control parameters include: $w_q$ =0.7 and C = 0.5.

Figure 4 illustrates FTP PDU loss probability for both TSP and the enhanced TSP scheme. Comparing the two, it can be seen that at 64 kbps FTP arrival rate their performance is identical i.e. very low PDU losses occur. But with higher FTP rates the enhanced TSP scheme is seen to give a much better loss performance with negligible PDU loss. Whereas for the basic TSP scheme, higher PDU loss probability is observed with higher FTP PDU arrival rates. These results indicate that significant improvement in non-real-time flow loss performance could be achieved for concurrent multimedia traffic, by incorporating threshold-based Iub flow control with the TSP buffer management scheme.

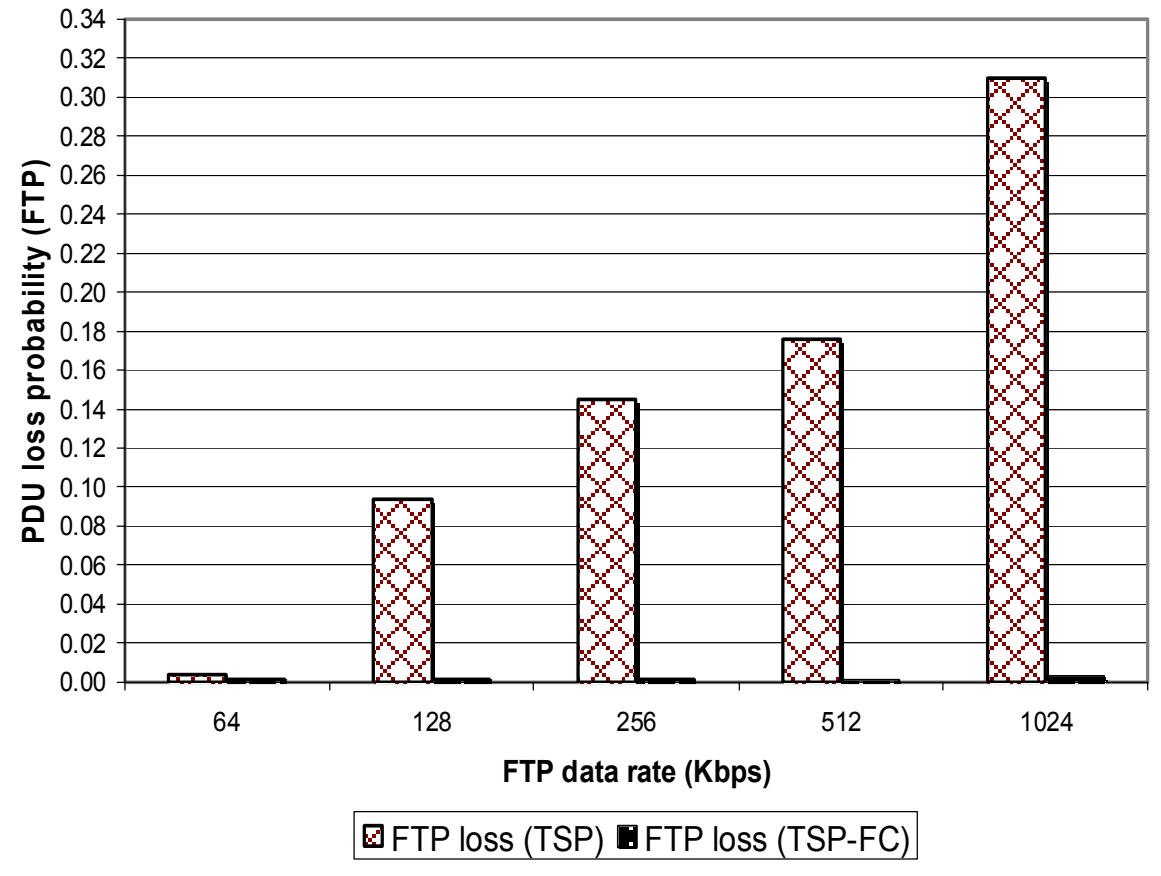

**Figure 4. FTP loss Performance of TSP and enhanced TSP schemes in a HSDPA Node B**

Figure 5 shows the throughput performance comparison between the TSP and the enhanced version. At all FTP packet arrival rates taken, the throughputs obtained were slightly higher than the arrival rates. This is because of the overhead of the PDU header added to segmented packets, which was taken into account in the throughput calculation. But as the results indicate identical performance for both schemes, it means that the enhanced version enabled better loss performance without sacrificing the throughput. Hence, air interface resource utilization in the enhanced TSP is quite high. Obviously, this depends on the buffer management and flow control parameters setting which must be chosen carefully to ensure optimum performance.

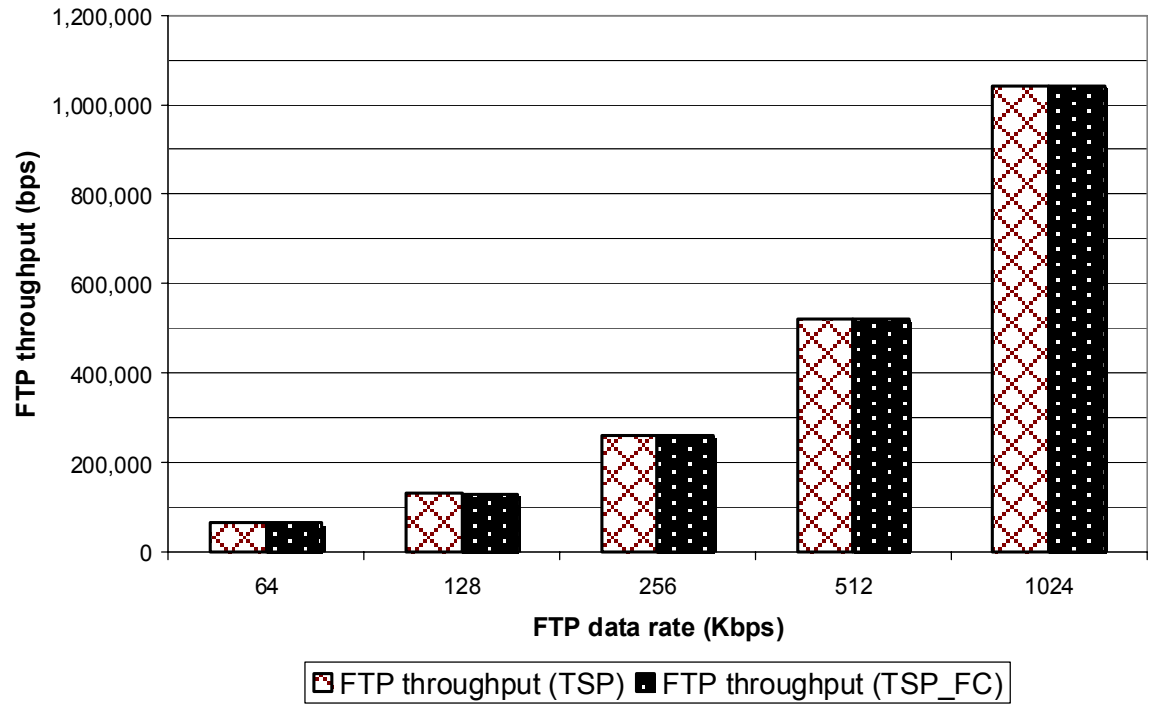

**Figure 5. FTP throughput Performance of TSP and enhanced TSP schemes in the Node B**

Mean Node B queuing delay of VoIP in the mixed multimedia traffic for both schemes is shown in figure 6. As expected, the delay VoIP performance is only slightly affected by the incorporation of the flow control mechanism. Recall that in TSP real-time flows in the multimedia are accorded service priority with precedence queuing. Furthermore, VoIP PDU transfer from RNC to Node B does not depend on Node B grant allocation. These account for the almost identical mean delay performance of both schemes.

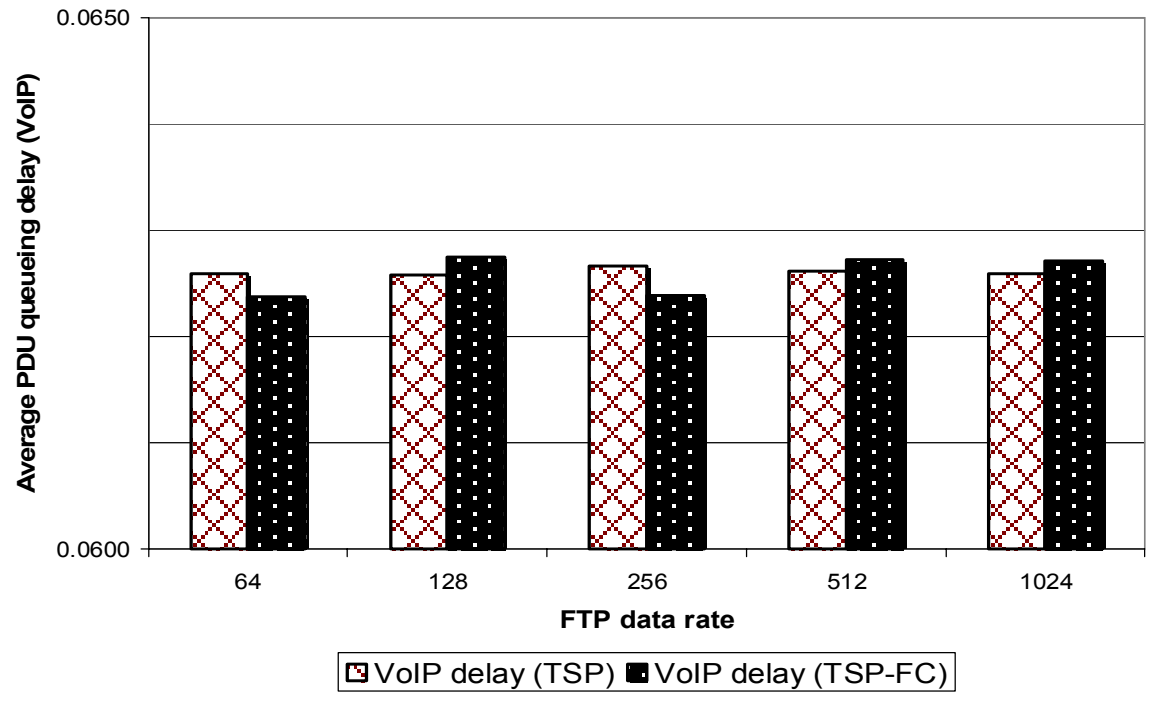

**Figure 6. VoIP queuing delay of TSP and enhanced TSP schemes in the Node B**

Figure 7 compares the PDU loss performance (or the blocking probability of arriving VoIP PDU to the Node B) of both schemes. Irrespective of the scheme, we see almost identical loss or blocking probability between 0.080 and 0.081. The performance is also insensitive to changes in FTP rates. The same reasons given earlier, explaining the results obtained in figure 6 apply here.

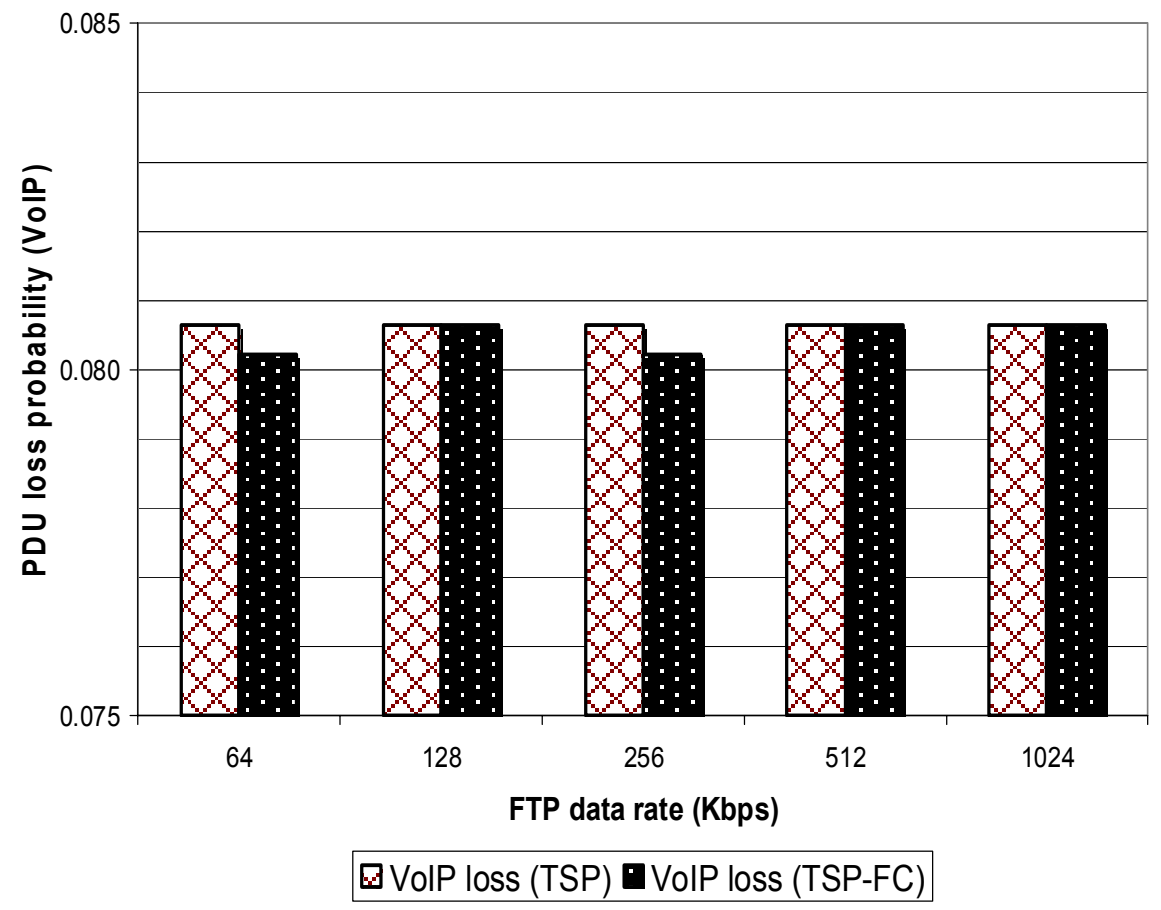

**Figure 7. VoIP loss Performance of TSP and the enhanced TSP Schemes in the Node B**

## 6. Concluding remarks

This paper has explored the extension of a previously proposed Time-Space priority buffer management scheme for QoS management of concurrent diverse flows in a HSDPA multimedia session. TSP is enhanced to include a possible Iub flow control strategy employing a threshold-based active queue management technique. By means of dynamic HSDPA simulation, the enhanced TSP is shown to dramatically improve loss performance of non-real-time flow in the multimedia stream, without compromising the QoS performance of the real-time flow. Thus, the enhanced TSP is a viable scheme that could improve the performance of HSDPA for broadband multimedia services support. The extension of TSP with other possible flow control mechanisms will be studied in future. Dynamic selection of thresholds to optimize performance will also be investigated.